# Microwave Nonlinearities of an Isolated Long $YBa_2Cu_3O_{7-\delta}$ Bi-crystal Grain Boundary

Sheng-Chiang Lee, Su-Young Lee, and Steven M. Anlage
Center for Superconductivity Research, Physics Department, University of Maryland, College Park, MD 20742-4111, USA

We measure the local harmonic generation from an $YBa_2Cu_3O_{7-\delta}$ (YBCO) bi-crystal grain boundary to examine the local Josephson nonlinearities. Spatially resolved images of second and third harmonic signals generated by the grain boundary are shown. The harmonic generation and the vortex dynamics along the grain boundary are modeled with the Extended Resistively Shunted Josephson (ERSJ) array model, which shows reasonable agreement with the experimental data. The model also gives qualitative insight into the vortex dynamics induced in the junction by the probing current distribution. A characteristic nonlinearity scaling current density $J_{NL} \sim 1.5\times10^5$ A/cm$^2$ for the Josephson nonlinearity is also extracted.

## Introduction

The nonlinear behavior of high-$T_c$ superconductors (HTSC) has been of great concern because it can potentially reveal the underlying physics of HTSC. While the microscopic origins of nonlinear response still remain uncertain, all superconductors have an intrinsic time-reversal symmetric nonlinearity associated with the Nonlinear Meissner Effect (NLME).[1,2,3] Calculations based on BCS theory and Ginzburg-Landau theory have been proposed to describe the harmonic generation (or intermodulation) response of the NLME.[4,5] Many experiments have been conducted to study 3rd order harmonic generation or intermodulation signals, which may arise from this intrinsic nonlinearity.[6,7,8,9]

On the other hand, extrinsic sources of nonlinearities dominate the nonlinear response of superconductors in most cases. They include grain boundaries, [10,11] enhanced edge currents,[12,13,14,15,16] and weakly coupled grains, [17,18,19,20,21,22,23,24]. A number of experiments have been conducted to understand and characterize the nonlinear properties of one of these sources by measuring artificially-prepared features, e.g. bi-crystal grain boundaries (GB).[17,19] Most of these experiments are done with resonant techniques, which by their nature study the averaged nonlinear response from the whole sample rather than locally.[25] Such techniques usually have difficulty in avoiding edge effects, which give undesired vortex entry due to the enhanced currents and defects along the etched edges,[16] and do not reveal the local intrinsic nonlinear properties of superconductors.

In our previous work, we have shown that by measuring the local nonlinear response in second and third harmonic generation, we can locally identifying the superconducting bi-crystal GB without the edge currents involved.[26] In this work, we present additional data and much more comprehensive analysis and simulations. The simulation results are examined to understand the vortex dynamics taking place in the middle of an infinite superconducting GB with AC currents passing through. In addition, a quantitative description of the nonlinear behavior of the GB is given in this work in terms of a nonlinear scaling current density.

## System Setup and Sample preparation

It has been established that harmonic generation and intermodulation measurement are the most sensitive methods to measure nonlinearities in superconductors.[27] In this paper, we employ the harmonic generation technique because it gives us access to both time-reversal symmetric and time-reversal symmetry-breaking sources of nonlinearity. As shown in Fig. 1, in order to detect the harmonic content generated on the surface of a sample, we send a single tone microwave signal at frequency *f*, which is low-pass filtered to guarantee the purity of the spectrum, to the sample via the coupling between a loop probe and the sample. The loop probe is made of a non-magnetic coaxial cable with its inner conductor forming a ~ 500 $\mu$m outer-diameter semi-circular loop shorted with the outer conductor. When this loop probe is close to a conducting surface, it couples to the surface magnetically, and induces microwave currents flowing on the surface with a geometry defined by the loop size, shape, and orientation. If there are any local nonlinear mechanisms present in the range of the induced microwave currents, additional microwave currents at multiples of *f* will be generated.[26,28]

We detect the lowest-order harmonic signals at 2*f* and 3*f* by measuring the microwave signal coupled back from the sample surface through the probe with the spectrum analyzer. Since the harmonic signals are extremely weak, they are amplified by ~ 60dB before entering the spectrum analyzer. High-pass filters are used before amplification to prevent amplifying the fundamental signal and getting a strong signal at frequency *f* into the spectrum analyzer. We measure the amplified second and third harmonic signals simultaneously with the spectrum analyzer as a function of temperature and position over the sample.

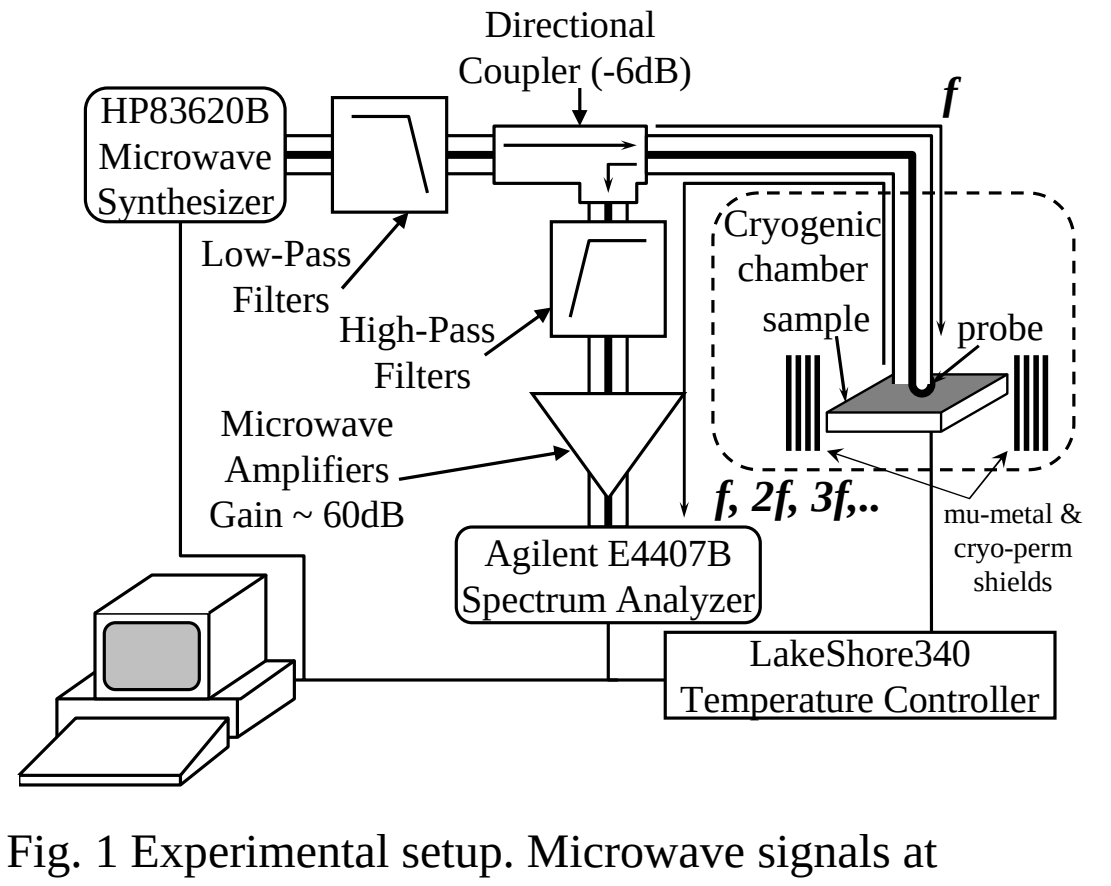

Fig. 1 Experimental setup. Microwave signals at frequency f $\cong$ 6.5 GHz are generated by the synthesizer, filtered and propagate to the sample via coaxial transmission lines. The loop probe creates RF currents on the surface of the sample. The sample creates response at 2*f* and 3*f*, and these signals couple back to the loop probe. The harmonic signals are measured by a spectrum analyzer after being filtered and amplified by ~ 60dB.

The sample is a 500Å-thick $YBa_2Cu_3O_{7-\delta}$ (YBCO) thin film deposited by pulsed laser deposition on a 30º misoriented bi-crystal $SrTiO_3$ substrate at nearly optimal doping level. The spatially averaged $T_c$ is 88.9K, as measured by ac susceptibility. The loop probe is placed 12.5µm above the sample, fixed by a Teflon™ sheet. The sample is kept in a high vacuum cryostat cooled with continuously flowing liquid Helium, and its temperature can be controlled between ~3.5 K and room temperature to within ± 10 mK. The probe can move in the x-, y-, and z-directions inside the cryostat to perform spatially resolved measurements. The sample is surrounded by two layers of mu-metal shielding, and two layers of cryo-perm shielding. They are supported by an ultra-low-carbon steel base to provide a low background magnetic field environment ($< 1\mu$G ) for the measurements. Thus all the measurements presented here are carried out in nominally zero dc magnetic field.

Measurements of temperature-dependent third-order harmonic power ($P_{3f}$) are first performed at locations above the grain boundary (GB) and away from the boundary (non-GB) as shown in the inset of Fig. 2. A strong peak in $P_{3f}(T)$ is observed around $T_c$ at all locations on the sample,[26,28,29] and these peaks show similar magnitudes as shown in Fig. 2.[30] This $P_{3f}$ peak near $T_c$ is predicted by all models of intrinsic nonlinearities of superconductors.[29] For example, the Ginzburg-Landau theory gives a reasonable fit to this peak, which is also shown in Fig. 2 as a solid line.[26,28] A small shift in temperature ~ 0.5K of this peak with location is also observed, which is most likely due to inhomogeneity at the grain boundary.

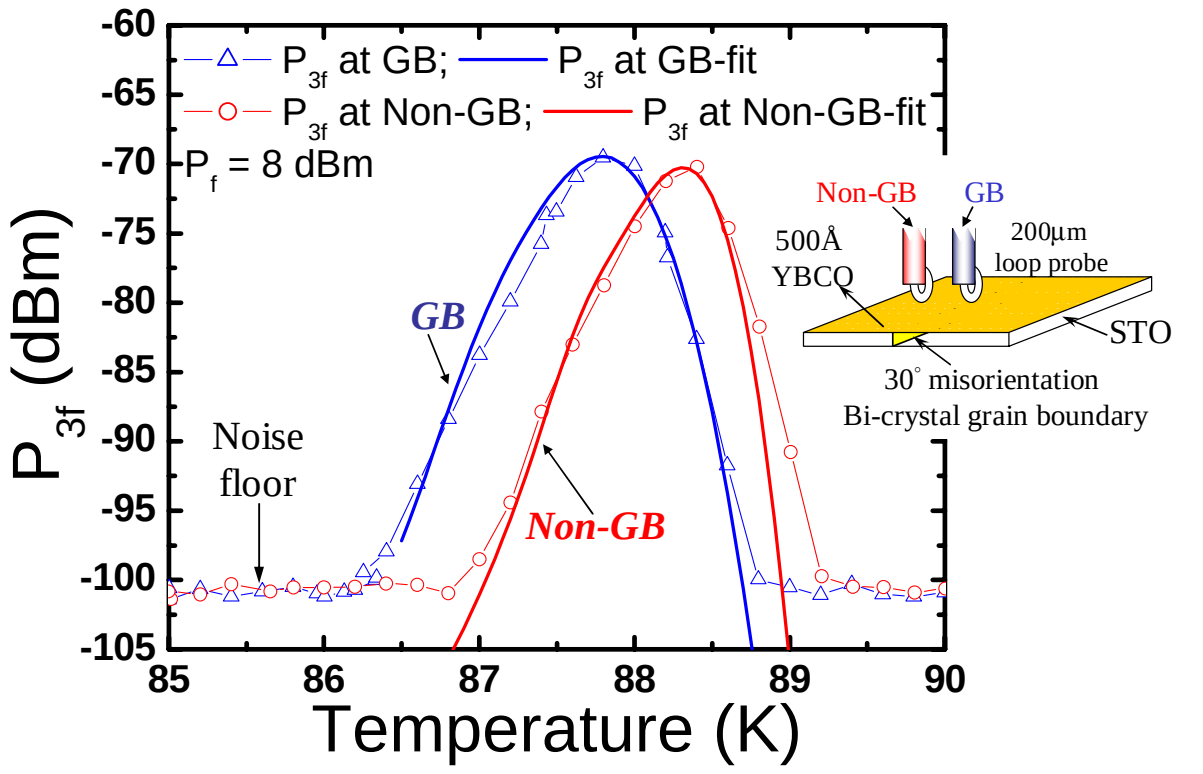

Fig. 2 $P_{3f}$(T) measured at GB (triangles) and non-GB (circles) locations near $T_c$ at f = 6.5 GHz. Model fitting based on the Ginzburg-Landau theory is also presented with $T_c$ = 88.5 K for non-GB and $T_c$ = 88.0 K for GB.

A significant difference in the third harmonic response with position is observed for $T < 0.9T_c$, where the GB shows significant enhancement ($P_{2f}$ ~ -70 dBm and $P_{3f}$ ~ -55 dBm) in both $P_{2f}$ and $P_{3f}$, but non-GB regions show only noise-level signals.[26] The enhanced $P_{2f}$ and $P_{3f}$ at the GB are due to the nonlinearity of the Josephson tunneling effect across the boundary, and our technique is capable of detecting and quantifying this local nonlinear mechanism, as discussed below.

## Data - Spatially Resolved 1D and 2D Images

To demonstrate that our microwave microscope is also able to spatially resolve the nonlinearity of the grain boundary, a measurement of $P_{2f}$ and $P_{3f}$ along a line crossing the grain boundary is performed at an intermediate temperature $T$ = 60K as shown in Fig. 3. A clear peak in both $P_{2f}$ and $P_{3f}$ is observed above the GB, with a width of about 1 mm, which is on the order of the size of the loop probe and its associated current distribution. This measurement is well interpreted and reproduced by the Extended Resistively Shunted Josephson junction model (ERSJ) discussed below.

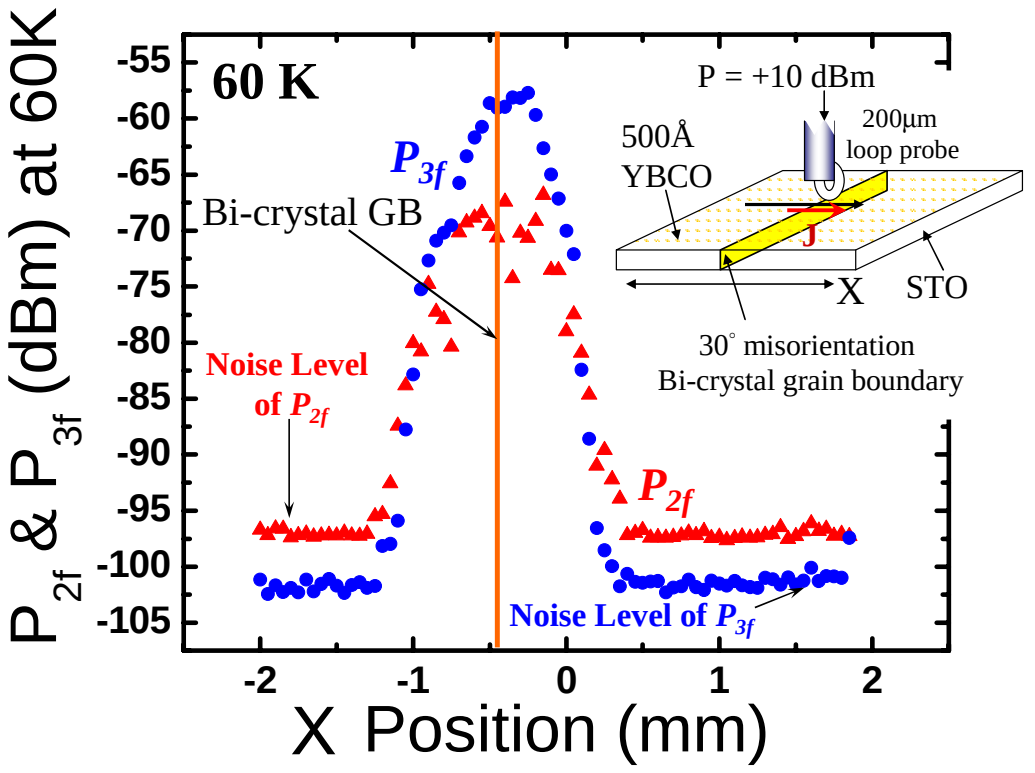

Fig. 3 $P_{2f}$ and $P_{3f}$ as a function of position across the bi-crystal grain boundary, taken at T = 60K, f = 6.5 GHz, and input power = +10 dBm. Enhancement of both $P_{2f}$ and $P_{3f}$ signals near the boundary is observed, with a width ~ 1 mm. This width is on the order of the spatial distribution of the current density, determined by the probe geometry.

To show that our microwave microscope is also capable of imaging non-uniformity of the bi-crystal grain boundary, we take a two-dimensional scanned image across the grain boundary with a loop probe at input microwave power of +10 dBm. The loop probe is oriented so that the currents are flowing across the grain boundary with a maximum current density $\sim 5\times10^4\ A/cm^2$. The scanning steps are $50\,\mu m$ across the boundary and $50\,\mu m$ along the boundary. As shown in Fig. 4, the bi-crystal grain boundary is identified in both the $P_{2f}$ and $P_{3f}$ images. It is clearly shown that the harmonic response due to the nonlinearities of the GB varies along the length of the grain boundary, and that the $P_{3f}$ image is more uniform than the $P_{2f}$ image. Also note that the probe was at least 3 mm away from all edges while taking this image, hence the contribution from the edge effect is avoided.

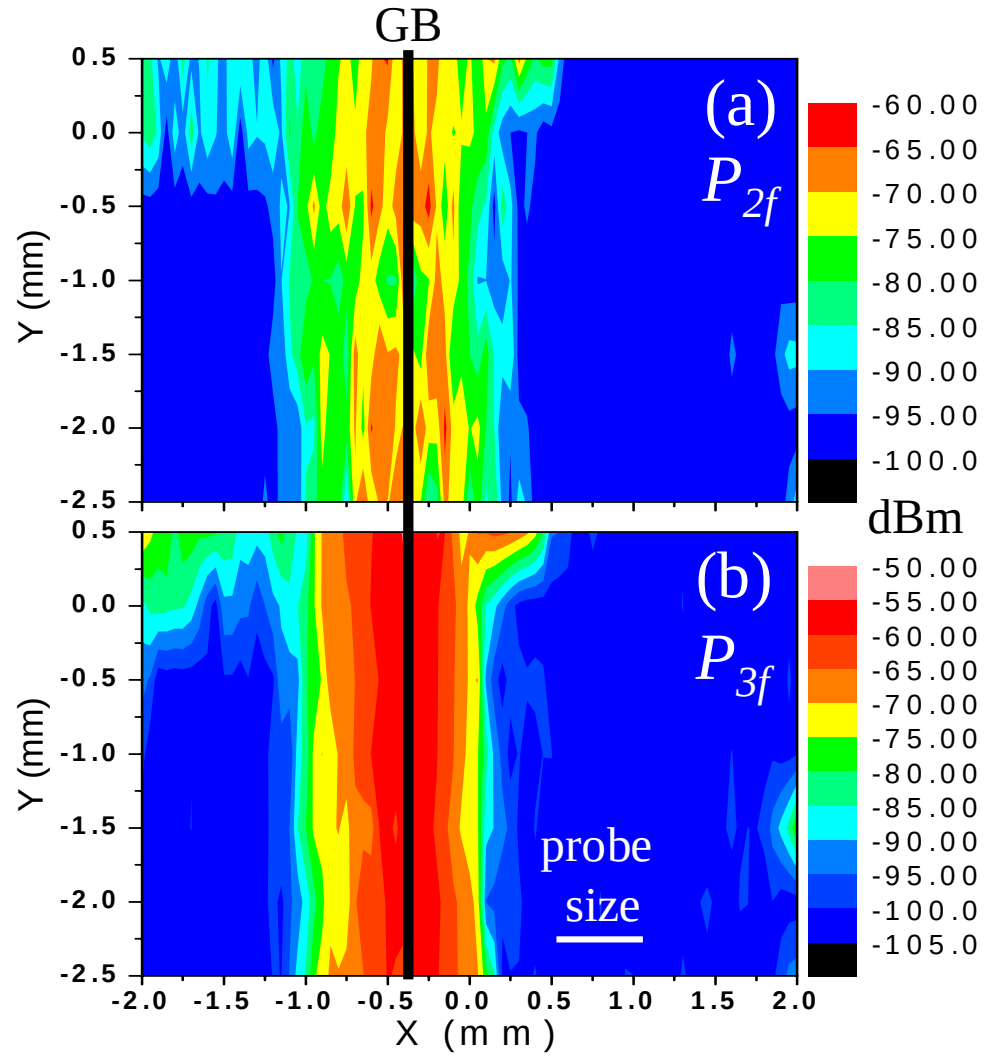

Fig. 4 Two dimensional images of $P_{2f}$ and $P_{3f}$ response on a $3\times4$ mm$^2$ area containing the grain boundary taken at T = 60K, f = 6.5 GHz, and input power + 12dBm. The RF currents are primarily sent in a direction perpendicular to the grain boundary. The GB is noted by the vertical black line, and is clearly identified by the enhancement of $P_{2f}$ and $P_{3f}$ signals above the background level.

## Extraction of the characteristic scales, $J_{NL}$ and $J_{NL}$'

There are many different microscopic models predicting various nonlinearities in superconductors. Most known models of nonlinear response can be expressed in a general form for the penetration depth $\lambda$ (or super-fluid density) assuming that time-reversal symmetry is preserved,

$$\frac{\lambda(T,J)^2}{\lambda(T,0)^2} = 1 + \left[\frac{J}{J_{NL}(T)}\right]^2 + \cdots, \qquad \text{Eq. 1}$$

where $J$ is the applied current density, and $J_{NL}(T)$ is a temperature-dependent, geometry-independent scaling current density, which can vary by orders of magnitude depending on the precise nonlinear mechanism and temperature (here we assume J << $J_{NL}$(T)). For example, the Ginzburg-Landau theory of the nonlinear Meissner effect yields $J_{NL} \sim 10^8\text{-}10^9$ $A/cm^2$, and a long 1D Josephson junction array expects $J_{NL}$ to be around $10^5$-$10^6$ $A/cm^2$ in the intermediate temperature region, $T/T_c \sim 0.5$.[31] To evaluate the ability of our microwave microscope to detect intrinsic superconducting nonlinearities due to different mechanisms, we extract the $J_{NL}$ from our data with the following algorithm.[28,32]

When a single-tone microwave signal at frequency $f$ is sent to a superconducting sample via the probe, microwave currents are induced in the sample due to the Meissner screening. The sample thickness is less than the magnetic penetration depth, and we assume that the super-currents flow uniformly over the thickness to screen out the magnetic field from penetrating through the sample. Since in this case most of the energy is carried by the super-currents, it is assumed that the kinetic inductance $L_{KI}$ of the superconductor dominates its nonlinear AC response. The kinetic inductance can be derived as[28,33]

$$L_{KI} = \mu_0 \int \frac{\iint_{\text{cross section}} \lambda^2(T,J) J^2 ds}{\left(\iint_{\text{cross section}} \vec{J}\cdot d\vec{s}\right)^2} dy, \qquad \text{Eq. 2}$$

where $\lambda$ is the magnetic penetration depth, $\vec{J} = \vec{J}(x,y)$ is the spatial distribution of the current density, and the numerator represents the energy stored in the super-currents. The denominator involves the total current flowing through a cross section perpendicular to the y-axis in the integral, as illustrated in Fig. 5. Using Eq. 1, Eq. 2 can be rewritten as

$$L_{KI} = \mu_0 \lambda(T,0)^2 \times$$

$$\int \frac{\iint_{\text{cross section}} J^2 ds + \iint_{\text{cross section}} \frac{J^4}{J_{NL}^2(T)} ds + \cdots}{\left(\iint_{\text{cross section}} \vec{J}\cdot d\vec{s}\right)^2} dy$$

$$\cong L_0 + \Delta L\, I^2, \qquad \text{Eq. 3}$$

where $L_0$ is the linear kinetic inductance and $\Delta L$ is the coefficient of the quadratic term in the nonlinear kinetic inductance.

Since the superconducting sample is driven by a microwave current, it is equivalent to an ac circuit with a driving ac current source $I_0 \sin(\omega t)$ and a lumped nonlinear inductor, and develops a voltage drop across the inductor, given by

$$V(t) \cong L_0 \frac{dI(t)}{dt} + \Delta L\, I(t)^2 \frac{dI(t)}{dt},$$

where $\omega$ is the driving frequency. By solving these equations, we can extract the harmonic content from the voltage solution, and the expected third harmonic power becomes,[28]

$$P_{3f} = \frac{|V_{3f}|^2}{2Z_0} = \frac{(\omega \Delta L I_0^{\,3}/4)^2}{2Z_0},$$

where $Z_0$ is the characteristic impedance of the coaxial transmission line, and

$$\Delta L = \frac{\mu_0 \lambda^2(T,0)}{I_0^{\,2} J_{NL}(T)^2} \int \frac{\iint_{\text{cross section}} J^4(x,y)ds}{\left(\iint_{\text{cross section}} \vec{J}\cdot d\vec{s}\right)^2} dy\ .$$

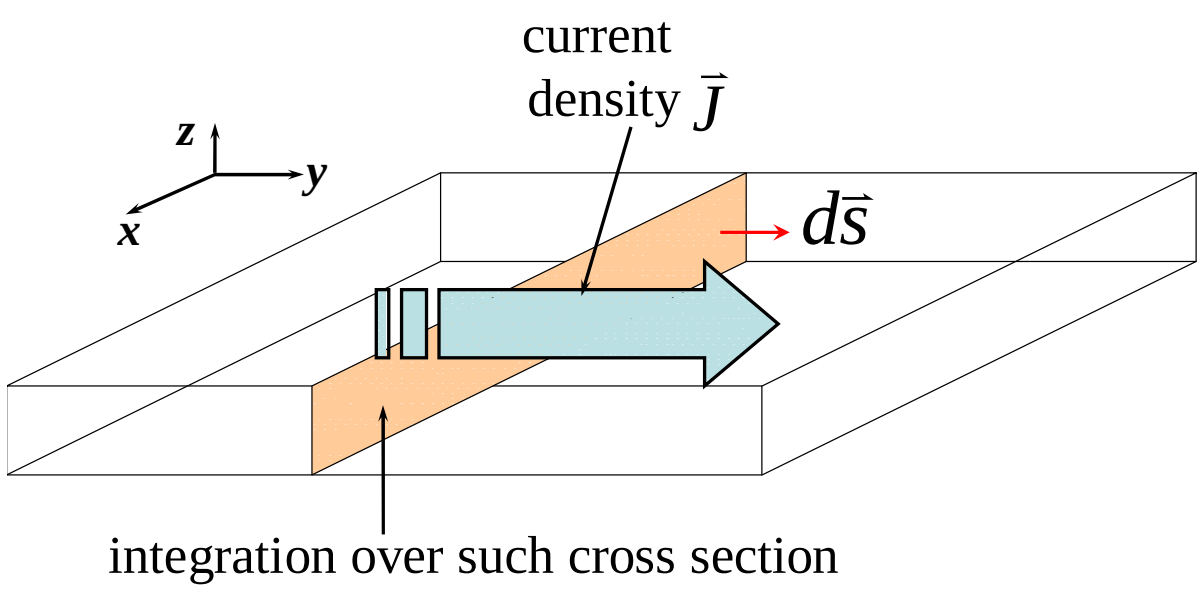

Fig. 5 Diagram of current flow and coordinate system used in the calculation of second and third harmonic power generated in the sample.

Since our sample is in the limit that its thickness $t$ is much less than the magnetic penetration depth, $t << \lambda$, the current density can be replaced by the surface current density, $\vec{J} = \vec{K}/t$, and the volume integral can be reduced to surface integral of $\vec{K} = \vec{K}(x,y)$,

$$\Delta L = \frac{\mu_0 \lambda(T,0)^2}{t^3 I_0^{\,2} J_{NL}(T)^2} \int \frac{\int K^4 dx}{\left(\int K_y dx\right)^2} dy,$$

so that

$$P_{3f} = \frac{1}{2Z_0}\left(\frac{\omega \mu_0 \lambda(T,0)^2}{4t^3 J_{NL}(T)^2}\right)^2 \left(I_0 \int \frac{\int K^4 dx}{\left(\int K_y dx\right)^2} dy\right)^2 \quad \text{.Eq. 4}$$

We use numerical solutions of Maxwell's equations (obtained using the High Frequency Structure Simulator (HFSS) software) to simulate the microwave currents induced on the sample surface by a loop probe, and perform a surface integral of $K^4$ and a line integral in the x-direction in Fig. 5 to obtain the figure of merit of the probe: $\Gamma \equiv I_0 \int \left| \int K^4 dx \Big/ \left(\int K_y dx\right)^2 \right| dy$. With an estimation of $\lambda(60K,0) = 2440$ Å and $\Gamma \approx 31.2$ $A^3/m^2$ at +12 dBm input power for our probe geometry, we can convert the measured $P_{3f}$ to $J_{NL}$ using Eq. 4. Recall that $J_{NL}$ is a geometry-independent quantity which indicates the responsible mechanism for the observed nonlinear responses.

To interpret the second harmonic data, we empirically introduce a term in Eq. 1 responsible for broken Time-Reversal symmetry, which is characterized by a TRSB characteristic nonlinear scaling current density $J_{NL}$',[28,29]

$$\frac{\lambda(T,J)^2}{\lambda(T,0)^2} = 1 + \left[\frac{J}{J_{NL}'}\right] + \left[\frac{J}{J_{NL}(T)}\right]^2 + \cdots, \qquad \text{Eq. 5}$$

Following the same algorithm, but now using the $P_{2f}$ data, we can develop a similar formula for extracting $J_{NL}$' from $P_{2f}$ data. Details of this algorithm can be found elsewhere.[28,29]

Figs. 6 and 7 show the spatially-resolved $J_{NL}$ and $J_{NL}$' across the bi-crystal grain boundary, obtained from the $P_{3f}$ and $P_{2f}$ data in Fig. 3. The $J_{NL}$ near the grain boundary is around $1.5\times10^5\ A/cm^2$, which is characteristic of a weak-link nonlinearity as we expect, and is in reasonable agreement with the work of Willemsen.[31] The $J_{NL}$' near the grain boundary (Fig. 7) is on the order of $10^7\ A/cm^2$, which is significantly larger than $J_{NL}$. The physical interpretation of $J_{NL}$' is not clear for the case of a bi-crystal grain boundary. The microscopic origins of $P_{2f}$ are discussed in detail below. The noise level in the current measurements limits the sensitivity of this setup to $J_{NL} < 2.1\times10^6\ A/cm^2$ and $J_{NL}' < 4.7\times10^8\ A/cm^2$.

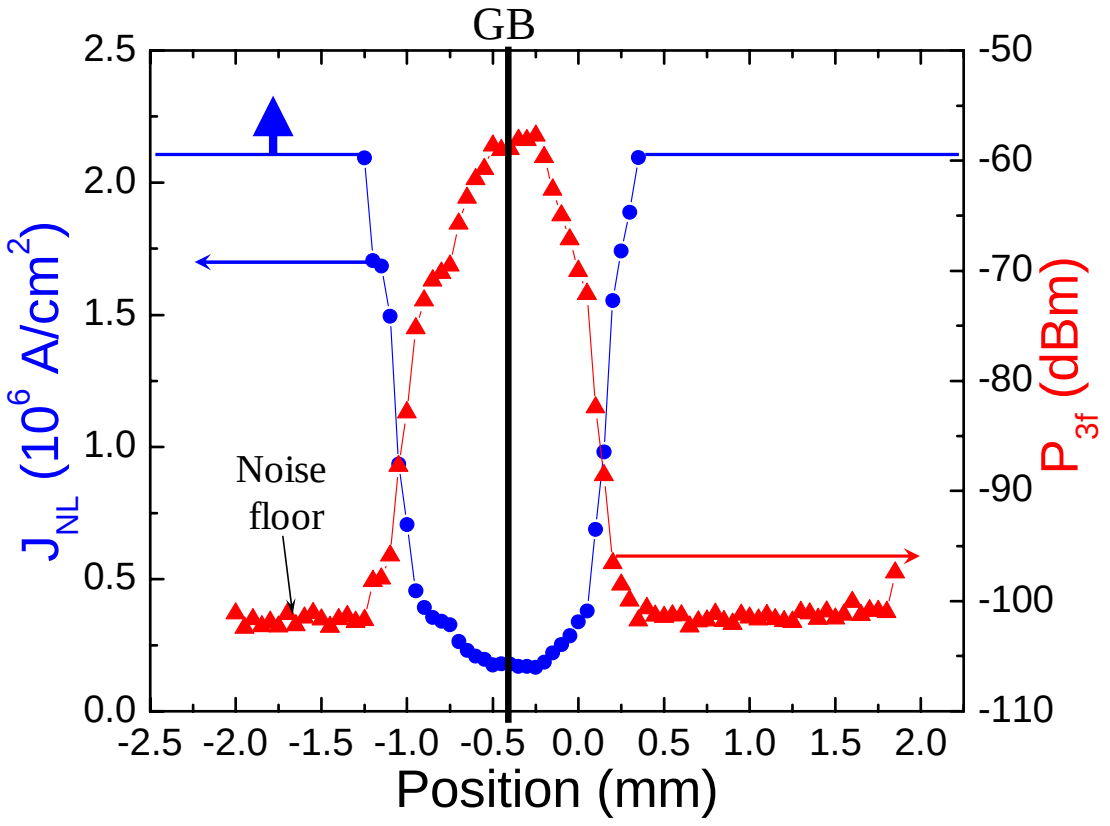

Fig. 6 The characteristic nonlinear scaling current density, $J_{NL}$ converted from the $P_{3f}$ data shown in Fig. 3 as a function of position. The $J_{NL}$ at the grain boundary is ~ $10^5$ $A/cm^2$, and this setup is limited to detecting nonlinearity with $J_{NL} < 2.1\times10^6\ A/cm^2$.

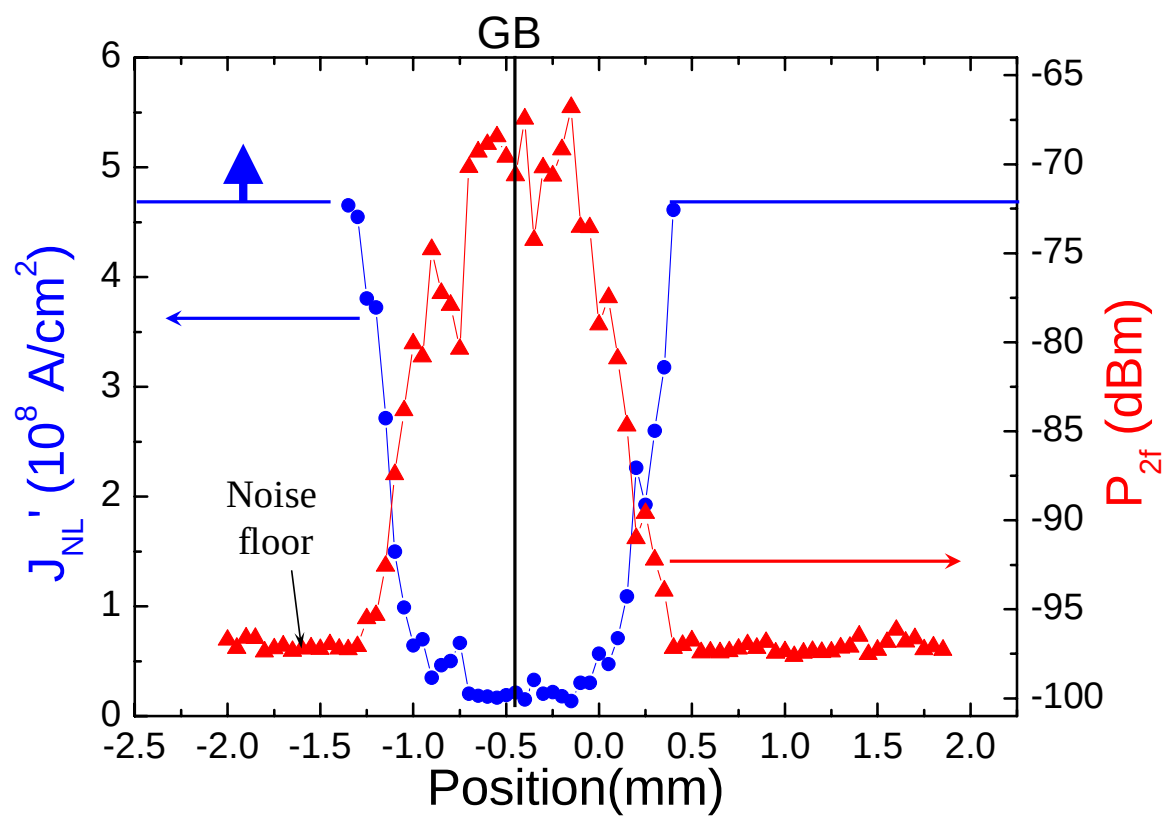

Fig. 7 The TRSB characteristic nonlinear scaling current density, $J_{NL}'$ converted from the $P_{2f}$ data of Fig. 3 as a function of position. The $J_{NL}'$ at the grain boundary is ~ $10^7$ A/cm$^2$, and the sensitivity is limited to nonlinearity with $J_{NL}' < 4.7\times10^8\ \mathrm{A/cm^2}$.

Better spatial resolution and sensitivity to larger $J_{NL}$ (corresponding to weaker nonlinearity) are desired. According to the algorithm sketched above, better sensitivity can be achieved by measuring thinner films (reducing *t*) and/or increasing the probe figure of merit $\Gamma$, while the spatial resolution can be improved by reducing the size of the loop probe and bringing it closer to the sample. We use both HFSS and an analytical model (described below and elsewhere[28]) to calculate $\Gamma$ for different sizes of probes, and find that the smaller the loop probe and the closer it gets to the sample, the larger the figure of merit $\Gamma$ (see Fig. 8). This means that better spatial resolution and higher sensitivity to $J_{NL}$ and $J_{NL}$' can be achieved simultaneously by using smaller loop probes.

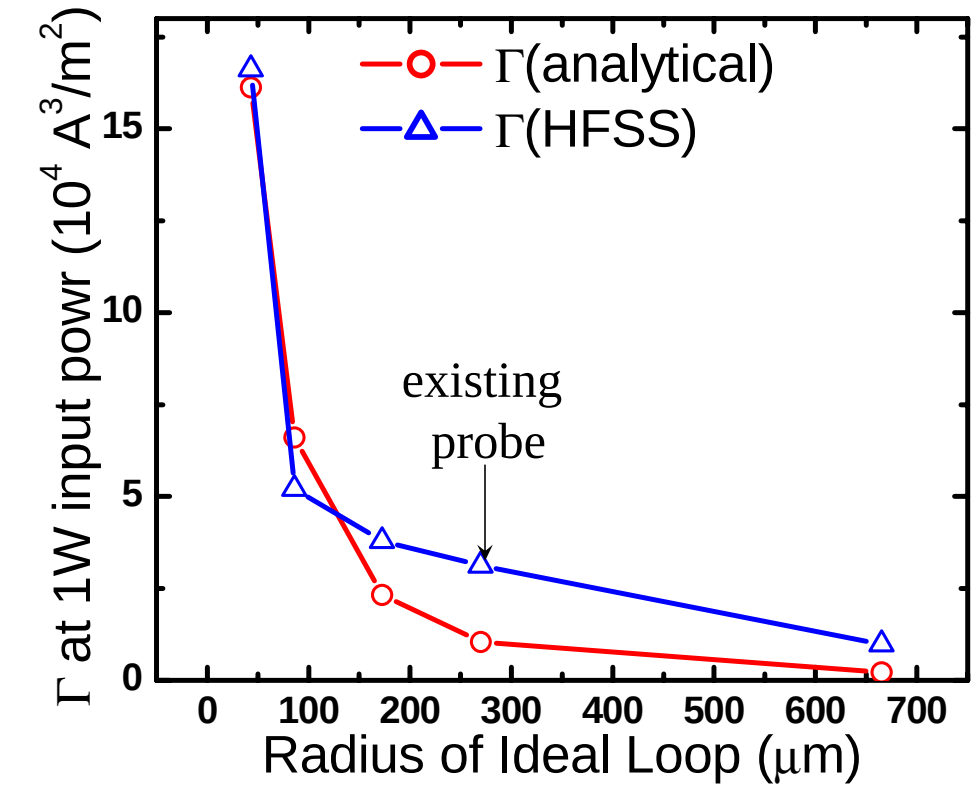

Fig. 8 The figure of merit $\Gamma$ evaluated for the loop geometry illustrated in Fig. 1 and 2, by analytical and numerical (HFSS) means as a function of loop probe size. The arrow indicates the $\Gamma$ of the existing probe. Both calculations suggest that smaller probes (which can be placed closer to the sample) will improve the sensitivity to weaker nonlinearities (larger $J_{NL}$).

## Model - Extended Resistively Shunted Josephson junction (ERSJ)

Our purpose is to develop a quantitative and microscopic understanding of the data presented in Figs. 3 and 4. It is well known that a purely single-tone AC current-biased single Josephson junction generates harmonics at all odd integer multiples of the driving frequency[10], which is measured in our data by $P_{3f}$. However, to obtain a more realistic and comprehensive understanding of a long weak-link junction, like the YBCO bi-crystal grain boundary, the Extended Resistively Shunted Josesphson (ERSJ) junction array model is introduced.

### Spatial Distribution of Harmonic Data

The ERSJ model we adopt was first used by Oates *et al.*[17,19] to describe the nonlinear behavior of a YBCO bi-crystal GB in their stripline superconducting microwave resonator.

In the ERSJ model, the long weak-link Josephson junction in the sample is modeled as a 1D array of 2001 single resistively shunted Josephson junctions combined in parallel, coupling with each other through lateral inductors estimated to be $l = 3\times10^{-11}\,\mathrm{H}$ (see Fig. 9).[28] Since the measurement takes place with a roughly 1 mm diameter current distribution around the center of a 10mm×10mm film, we have an essentially infinite 2D superconducting plane with no edge effect involved. To simulate this, the array is terminated by large lateral coupling inductors with $l = 10^{-8}\ H$ before the last junctions. The spacing between junctions is determined by the Josephson penetration depth. From prior work with these junctions, we estimate each junction in the ERSJ model to have a Josephson penetration depth $\lambda_J \sim 1\,\mu\mathrm{m}$, a critical current of $6\,\mu\mathrm{A}$, and a shunt resistance R of $50\,\Omega$. We assume that all the junctions, resistors, and inductors are identical, and the junctions are equally spaced. We use a program, WRSpice™, obtained from Whiteley Research Inc. to carry out the simulations of this model. The physical quantities discussed above are parameters that can vary in the WRSpice™ simulation.

It is assumed that the biasing microwave current with frequency = 6.5 GHz applied to each junction ($I_n$) in the model varies according to the surface current distribution on the film induced by the loop probe. The current distribution is estimated from two calculations. First is a simplified analytical model of an ideal circular loop in a vertical plane, with radius $270\,\mu\mathrm{m}$, coupling to a perfectly conducting horizontal plane $382.5\,\mu\mathrm{m}$ away from the center of the loop.[28] The magnitude of the current density is determined by a much more sophisticated microwave simulation software, HFSS by Ansoft™. The full geometry of the loop probe is used in this calculation, and it also produces a similar current distribution. The maximum current density on the sample is ~ $5\times10^{4}\,\mathrm{A/cm^{2}}$ for +12 dBm input power. A cross section through the peak current distribution is shown in Fig. 9.

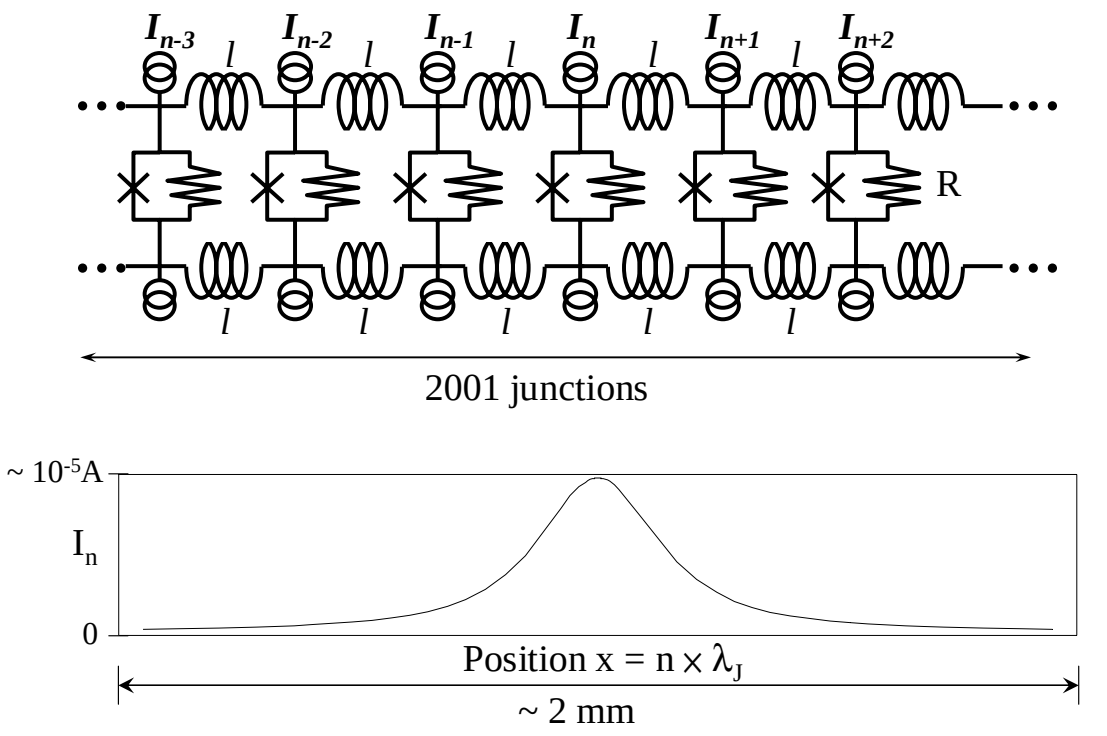

Fig. 9 Schematic diagram of the ERSJ model. The crosses are Josephson junctions. The lower panels shows a typical current distribution applied to the Josephson junction array for +12 dBm input power.

The calculated two dimensional current distribution is applied to each junction in the WRSpice™ program. To reproduce the spatial distribution of the measured $P_{2f}$ and $P_{3f}$ in Fig. 10 from the model, we take a one dimensional slice from the two-dimensional current distribution at 101 locations along the scanning direction and apply it to the junctions in the ERSJ model. We sum up the nonlinear potential differences across all junctions, calculated by WRSpice™. The simulations are done in transient analysis, which simulates the evolution of the system in time. The potential differences are found to be periodic a few periods after starting the numerical analysis. We average the potential differences between the 5th period and 65th period to reduce the numerical error from the calculation, and extract the higher harmonics from this collective nonlinear potential difference via Fourier transformation. The calculated $P_{2f}$ and $P_{3f}$ show good quantitative agreement with experimental results in both magnitude and spatial resolution as shown in Fig. 10.

Since a single pure-tone AC current-biased Josephson junction only generates $P_{3f}$ signal (as well as other odd harmonics), the $P_{2f}$ signals are attributed to the presence and motion of Josephson vortices generated along the long grain boundary. The voltages across the junctions become time-irreversible because of the motion of the vortices, and therefore contain second harmonic content. This interpretation is confirmed by simulating the ERSJ model with no lateral coupling inductors (un-coupled ERSJ), with each junction responding to its biasing current independently. The calculated $P_{3f}$ of the un-coupled ERSJ model has larger magnitude and narrower spatial distribution due to the lack of lateral coupling, and no $P_{2f}$ is generated in the uncoupled case.[26] The strong presence of $P_{2f}$ in the experimental data is proof that the collective behavior of the Josephson system is crucial to our understanding.

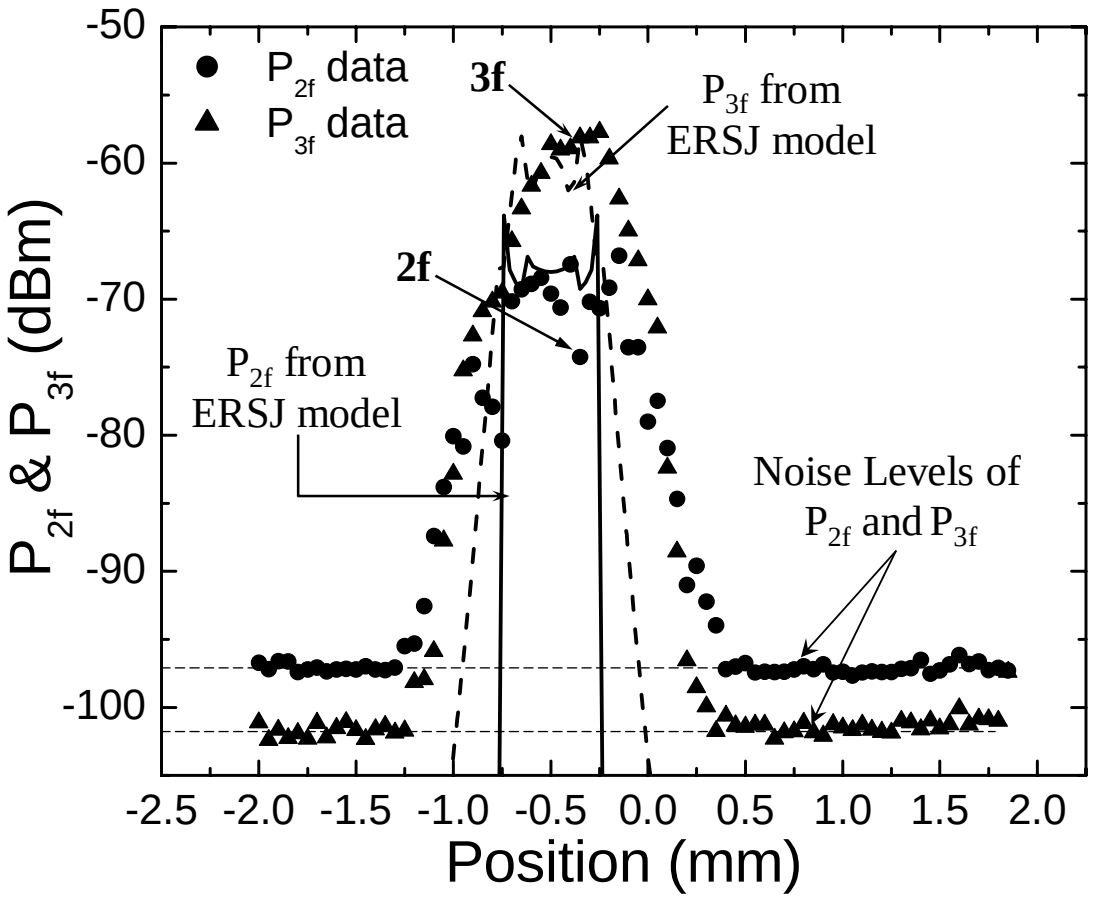

Fig. 10 $P_{2f}$(X) (dashed line) and $P_{3f}$(X) (solid line) simulated by the ERSJ model are compared with the experimental $P_{2f}$ (circles) and $P_{3f}$ ( triangles) from Fig. 3.

The simulated $P_{2f}$ and $P_{3f}$ by the ERSJ model are shown in Fig. 10 with the experimental results. It appears that the model well describes the magnitudes and spatial distribution of the $P_{2f}$ and $P_{3f}$ data. However, the details of the data are not well described by the model because the non-uniformity in the sample (as seen in Fig. 4) is not present in the model. The broader distribution of $P_{2f}$ and $P_{3f}$ data in Fig. 10 may be due to inhomogeneity and pinning sites in the grain boundary, which are not included in the model. The wiggles shown in the calculated solid and dashed lines in Fig. 10 are due to the numerical nature of the calculation. They are reduced tremendously by averaging over more periods, at the expense of a significant increase of computation time and required computation resources. Because of these constraints, the results presented in this paper are averaged over only 60 periods.

It is also noted that by applying strong enough microwave power, the Josephson array system becomes chaotic.[10] This phenomenon is also seen in our simulations at much higher input powers ( > 30 dBm), but does not play a role at the powers used in the experiments (~ 10 dBm).

**Vortex Dynamics**

To further our fundamental understanding of the physics governing the local nonlinearities, especially the $P_{2f}$ responses, we use the ERSJ model to evaluate the nucleation and motion of Josephson vortices in the middle of a driven infinite superconducting GB.

A long Josephson junction can be described by the sine-Gordon equation,[34, 35]

$$\lambda_J^2 \frac{\partial^2 \Delta\gamma(x,t)}{\partial x^2} = \sin\Delta\gamma(x,t) + \frac{L_J}{R_J}\frac{\partial \Delta\gamma(x,t)}{\partial t} + C_J L_J \frac{\partial^2 \Delta\gamma(x,t)}{\partial t^2},$$

where $\Delta\gamma$ is the gauge-invariant phase difference across the junction, $\lambda_J$ is the Josephson penetration depth, $L_J \equiv \Phi_0/2\pi J_c$, $R_J \equiv \rho d$, $\rho$ and d are the junction resistivity and thickness, and $C_J = \varepsilon/d$. $L_J$, $R_J$, and $C_J$ are all quantities per unit length. Since our ERSJ model is equivalent to solving this equation on a discrete one-dimensional lattice, we calculate the key quantity $\Delta\gamma(n,t)$, where n indicates the $n^{th}$ junction, to extract other physical quantities, such as the current, magnetic field, and flux at each junction.

The magnetic field along the grain boundary is given by,

$$B(x) = \frac{\Phi_0}{2\pi d_m}\frac{\partial \Delta\gamma(x,t)}{\partial x},$$

where $\Phi_0$ is the flux quantum, and $d_m = d + 2\lambda \coth(t/\lambda) \cong 2\lambda\coth(t/\lambda)$ is the magnetic thickness of the junction.[35] Since the distance between the junctions is $\lambda_J$ in the model, the flux between adjacent junctions is determined by

$$\Phi(n) = B(n)\times(d_m\cdot\lambda_J) = \frac{\Phi_0\lambda_J}{2\pi}\left.\frac{\partial\Delta\gamma(x,t)}{\partial x}\right|_{x=n\times\lambda_J}.$$

From the WRSpice™ simulation, we obtain $\Delta\gamma$, the current flowing through each junction, and the voltage across each junction as functions of space and time. Therefore, by identifying the cores of vortices, we can map out the trajectories of vortices in a space-time plot. The cores of vortices are identified by finding where $\Delta\gamma$ is an odd multiple of $\pi$. The vortex trajectories simulated at different microwave input powers are shown in Fig. 11 (one vortex-anti-vortex (VAV) pair in a RF cycle) and Fig. 12 (2 VAV pairs in an RF cycle). The calculations assume that the loop probe is located directly on top of the junction.

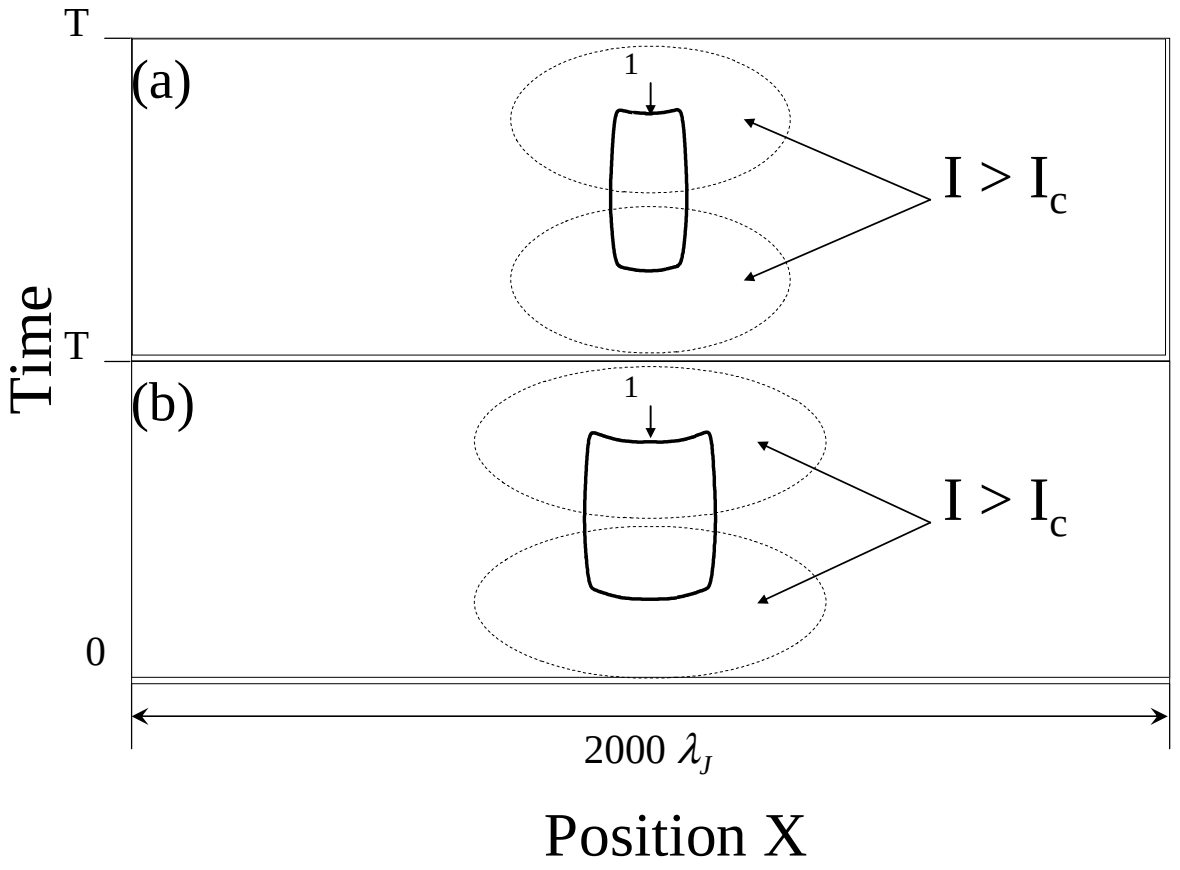

Fig. 11 Vortex trajectories in the ERSJ model simulated at different input microwave powers (+13 and +17 dBm). The power increases from (a) to (b). The vertical axis is time, which progresses upward from t = 0 to t = T ~ 154 ps (RF period) in each graph. The harmonic generation from each simulation are (a) $P_{2f}$ = -67 dBm and $P_{3f}$ = -72 dBm, and (b) $P_{2f}$ = -70 dBm and $P_{3f}$ = - 59 dBm.

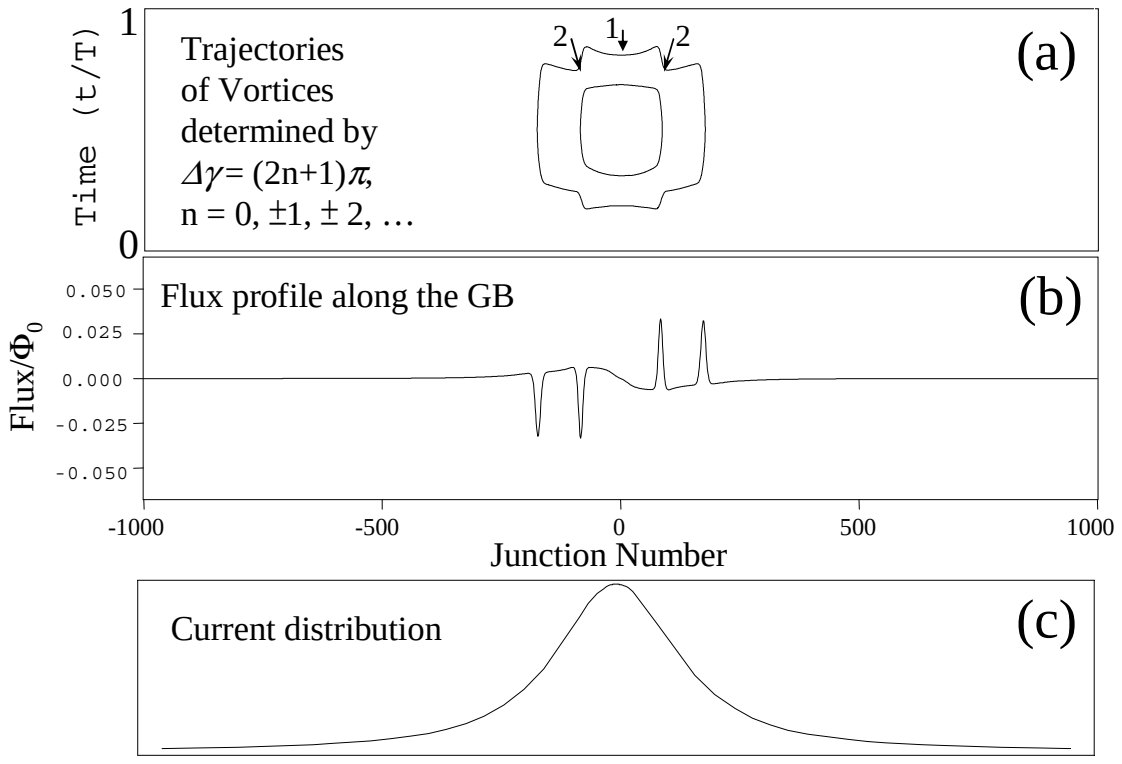

Fig. 12 (a) Vortex trajectories simulated at +21 dBm, which generates 2 VAV pairs in one RF cycle. (b) The flux profile along the grain boundary is also plotted at t = 0.5 T for comparison, along with (c) the applied RF current distribution along the grain boundary. The harmonic generation from this simulation are $P_{2f}$ = -68 dBm and $P_{3f}$ = -60 dBm.

The slopes of the trajectories in Figs. 11 and 12 represent the inverse of the speed of a vortex. Note that time proceeds upward in these plots. If the trajectory is vertical in the plot, the vortex is stationary. If the trajectory is horizontal, the vortex is moving very fast. It is noted that the simulation does not demonstrate smooth motion of vortices. When a VAV pair is first nucleated at the center of the junction in the first half of a RF cycle (Fig. 11), the vortices are expelled very rapidly from the center to locations that are far apart (>> $\lambda_J$), where they remain for most of their existence. We note that in Fig. 11 (a) and (b) when the rf power is increased, the pairs move farther apart. Also note that the original VAV pair is annihilated by a second anti-parralleVAV pair nucleated in the second half of the RF cycle. As the power is further increased, the vortices are expelled further and further away, and eventually additional VAV pairs are generated in the first half of a RF cycle (Fig. 12).

We interpret the discrete motion of vortices (seen in Fig. 12(a)) as a result of the simultaneous breakdown of many neighboring junctions. Considering the current distribution that we apply to the junctions in the ERSJ model as shown in Fig. 9, many junctions can experience currents exceeding their critical current and break down together. This is because the length scale of the current distribution variation is set by the loop size, which is ~ $10^3 \lambda_J$ . Since the junctions are experiencing AC currents at GHz frequencies, they may not break down spontaneously when they experience currents exceeding $I_c$ (Josephson critical current). The ovals in Fig.11 mark the times and junctions which experience $I > I_c$. These simulations indicate that junctions break down only when they experience $I > I_c$ over a significant portion of time in each period.

As the currents reverse direction in the second half of a RF cycle, another VAV pair with opposite polarity to the pair in the first half of a RF cycle is nucleated at the center of the junction. These new vortices rapidly move out to the nearly stationary locations of the previous pair and annihilate them (as pointed out by arrow 1 in Fig. 11 and 12). As the power is increased, the VAV pairs in the second half of a RF cycle can even be nucleated near stationary locations and annihilate the vortex at the next stationary location. (as pointed out by arrow 2 in Fig. 12)

The flux profile along the junction at t = T/2 is also shown with the trajectories in Fig. 12 (b). The spikes in the flux profile correspond to the stationary locations of the vortices. However, the locations of the spikes are fixed throughout the whole RF cycle, and do not move with the vortex trajectories. It is noted that only when a vortex moves to one of the locations does the corresponding spike contain one integer flux quantum. Otherwise, the total flux beneath the spike is less than one flux quantum. Also note that the net flux in the junction remains zero at all times.

In summary, the simulation shows that nucleation of VAV pairs takes place in the first half of the RF cycle and annihilation in the second half of the RF cycle. The discrete motion of the VAV pairs is likely due to the simultaneous break-down of many junctions which experiences $I > I_c$ over a significant portion of a RF period. The VAV pairs are nominally nucleated at the center of the junction (directly beneath the probe). As the input microwave power to the junction increases, more VAV pairs are nucleated and pushed away from the center in the first half of RF cycle. In this case, VAV pairs are annihilated by another set of VAV pairs nucleated in the second half of the RF cycle at locations next to each vortex and anti-vortex. From the time-reversal asymmetries of the vortex trajectories, we find that this nucleation and annihilation process (vortex motion) is responsible for the observed second harmonic generation.

## Conclusion

We have demonstrated local measurement of Josephson nonlinearities on a YBCO bi-crystal grain boundary. Spatially resolved harmonic generation images of this boundary are shown, and imply the ability of our microscope to measure non-uniformity along the boundary. Harmonic generation from the grain boundary is simulated by the ERSJ model, which agrees reasonably well with experimental data in the aspects of the magnitude and spatial distribution of both second and third harmonics. Vortex dynamics is also evaluated through this model. We also extract the nonlinear scaling current density $J_{NL}$ for the Josephson nonlinearity in our sample, and find $J_{NL} \sim 1.5\times10^5$ A/cm$^2$ at 60 K, which is reasonable and similar to that found by others.

We gratefully acknowledge the technical support of Dr. Steve Whiteley in performing simulations using WRSpice™, help from Dragos Mircea in taking harmonic generation data, and discussion with Dr. C. J. Lobb. We also acknowledge the support from NSF/GOALI DMR-0201261, and the UMD/Rutgers NSF-MRSEC DMR-00-80008 through the Microwave Microscope Shared Experimental Facility.

[1] S. K. Yip and J. A. Sauls, Phys. Rev. Lett. **69**, 2264 (1992).
[2] D. Xu, S. K. Yip, and J. A. Sauls, Phys. Rev. B **51**, 16233 (1995).
[3] J. Gittleman, B. Rosenblum, T. E. Seidel, and A. W. Wicklund, Phys. Rev. **137**, A527 (1965).
[4] T. Dahm and D. J. Scalapino, J. Appl. Phys. **81**, 2002 (1997).
[5] T. Dahm and D. J. Scalapino, Phys. Rev. B **60**, 13125 (1999).
[6] T. B. Samoilova, Supercond. Sci. Technol. **8** 259 (1995).
[7] Charles Wilker, Zhi-Yuan Shen, Philip Pang, William L. Holstein, and Dean W. Face, IEEE Trans. Appl. Supercond. **5**, 1665 (1995).
[8] G. Hampel, B. Batlogg, K. Krishana, N. P. Ong, W. Prusseit, H. Kinder, and A. C. Anderson, Appl. Phys. Lett. **71**, 3904 (1997).

[9] G. Benz, S. Wünsch, T.A. Scherer, M. Neuhaus, and W. Jutzi, Physica C **356**, 122 (2001).
[10] L. Ji, L. Ji, R. H. Sohn, G. C. Spalding, C. J. Lobb, and M. Tinkham, Phys. Rev. B **40**, 10936 (1989).
[11] J. McDonald and John R. Clem, Phys. Rev. B **56**, 14723 (1997).
[12] T. Dalm and D. J. Scalapino, J. Appl. Phys. **82**, 464 (1997).
[13] R. B. Hammond, E. R. Soares, B. A. Willemsen, T. Dahm, D. J. Scalapino, and J. R. Schrieffer, J. Appl. Phys. **84**, 5662 (1998).
[14] O. G. Vendik, I. B. Vendik, and D. I. Kaparkov, IEEE Trans. Microwave Theory Tech. **46**, 469 (1998).
[15] Y. Mawatari and J. R. Clem, Phys. Rev. Lett. **86**, 2870 (2001).
[16] A. P. Zhuravel, A. V. Ustinov, H. Harshavardhan, and S. M. Anlage, Appl. Phys. Lett. **82**, 4979 (2002).
[17] D. E. Oates, P. P. Nguyen, Y. Habib, G. Dresselhaus, M. S. Dresselhaus, G. Koren, and E. Polturak, Appl. Phys. Lett. **68**, 705 (1996).
[18] J. McDonald, J. R. Clem, and D. E. Oates, J. Appl. Phys. **83**, 5307 (1998).
[19] D. E. Oates, Y. M. Habib, C. J. Lehner, L. R. Vale, R. H. Ono, G. Dresselhaus, and M. S. Dresselhaus, IEEE Trans. Appl. Supercond. **9**, 2446 (1999).
[20] H. Hoshizaki, N. Sakakibara, and Y. Ueno, J. Appl. Phys. **86**, 5788 (1999).
[21] L. Hao, J. C. Gallop, A. J. Purnell, and L. F. Cohen, J. Super. **14**, 29 (2001).
[22] J. C. Booth, L. R. Vale, R. H. Ono, and J. H. Claassen, J. Super. **14**, 65 (2001).
[23] J. Halbritter, Supercond. Sci. Technol. **16**, R47 (2003).
[24] M. A. Hein, R. G. Humphreys, P. J. Hirst, S. H. Park, and D. E. Oates, J. Super. **16**, 895 (2003).
[25] W. Hu, A. S. Thanawalla, B. J. Feenstra, F. C. Wellstood, and S. M. Anlage, Appl. Phys. Lett. **75**, 2824 (1999).
[26] Sheng-Chiang Lee and Steven M. Anlage, Appl. Phys. Lett. **82**, 1893 (2003).
[27] H. Xin, D. E. Oates, G. Dresselhaus, and M. S. Dresselhaus, Phys. Rev. B **65**, 214533 (2002).
[28] Sheng-Chiang Lee, Ph.D. Dissertation, University of Maryland (2004).
[29] Sheng-Chiang Lee, Mathew Sullivan, Gregory R. Ruchti, Steven M. Anlage, Benjamin Palmer, B. Maiorov, and E. Osquiguil, submitted to Phys. Rev. B (2004), cond-mat/0405595.
[30] The fact that the transition occurs near the Tc determined by ac susceptibility implies that the probe does not increase the temperature of the sample significantly. Other discussion ruling out systematic errors are presented in Refs. [26], [28], and [29].
[31] Balam A. Willemsen, K. E. Kihlstrom. T. Dahm, D. J. Scalapino, B. Gowe, D. A. Bonn, and W. N. Hardy, Phys. Rev. B **58**, 6650 (1998).
[32] James C. Booth, J. A. Beall, D. A. Rudman, L. R. Vale, and R. H. Ono, J. Appl. Phys. **86**, 1020 (1999).
[33] Nathan Bluzer and David K. Fork, IEEE Trans. Magn. **28**, 2051 (1992).
[34] C. J. Lehner, D. E. Oates, Y. M. Habib, G. Dresselhaus, and M. S. Dresselhaus, J. Supercond. **12**, 363 (1999).
[35] E. Goldobin, A. M. Klushin, M. Siegel, and N. Klein, J. Appl. Phys. **92**, 3239 (2002).